\documentclass{ws-procs975x65}

\newcommand{\beq}{\begin{equation}}
\newcommand{\eeq}{\end{equation}}
\newcommand{\bea}{\begin{eqnarray}}
\newcommand{\eea}{\end{eqnarray}}

\begin{document}

\title{Gauge-Higgs Unification at the LHC}

\author{Nobuhito Maru\footnote{Speaker of the conference.}}

\address{Department of Physics, Chuo University,\\
Tokyo, 112-8551, Japan \\
E-mail: maru@phys.chuo-u.ac.jp}

\author{Nobuchika Okada}

\address{Department of Physics and Astronomy, University of Alabama,\\
Tuscaloosa, AL 35487, USA\\
E-mail: okadan@ua.edu}

\begin{abstract}
Higgs boson production by the gluon fusion and 
 its decay into two photons at the LHC 
 are investigated in the context of the gauge-Higgs unification scenario. 
The qualitative behaviors for these processes 
 in the scenario 
 are quite distinguishable from those of the Standard Model 
 and the universal extra dimension scenario 
 because of the overall sign difference for the effective couplings 
 induced by one-loop corrections through Kaluza-Klein (KK) modes. 
\end{abstract}

\keywords{Gauge-Higgs unification, Higgs production and decay at LHC}

\bodymatter

\section{Introduction}

Gauge-Higgs unification (GHU) is a fascinating scenario solving the hierarchy problem without invoking supersymmetry \cite{gh}. 
In this scenario, Higgs scalar field in the Standard Model (SM) is identified with 
 extra components of higher dimensional gauge field. 
The remarkable thing in this scenario is that the quantum correction to Higgs mass is finite 
due to the higher dimensional gauge symmetry 
regardless of the nonrenormalizability of the theory \cite{HIL} (See also \cite{others}). 
Such a UV insensitivity in other physical observables has been also investigated 
for S and T parameters \cite{LM1}, g-2 \cite{ALM}, the violation of gauge-Yukawa universality \cite{LM2} 
and the gluon fusion (two photon decay) of Higgs boson \cite{Maru}. 
The last one is the issue discussed in this talk. 

The Large Hadron Collider (LHC) started its operation again 
 and the collider signatures of various new physics models 
 beyond the SM have been extensively studied. 
The GHU shares the similar structure with the universal extra dimension (UED) scenario \cite{UED}, 
 namely, Kaluza-Klein (KK) states of the SM particles appear. 
The collider phenomenology on the KK particles will be 
 quite similar to the one in the UED scenario. 
A crucial difference should lie in the Higgs sector, 
 because the Higgs doublet originates from the higher dimensional gauge field 
 and its interactions are controlled by the higher dimensional gauge invariance. 
The discovery of Higgs boson is expected at the LHC, by which 
 the origin of the electroweak symmetry breaking 
 and the mechanism responsible for generating fermion masses 
 will be revealed. 
Precise measurements of Higgs boson properties will 
 provide us the information of a new physics relevant to the Higgs sector

In this talk, we investigate the effect of GHU on Higgs boson phenomenology at the LHC, 
 namely, the production and decay processes of Higgs boson \cite{MO} (See also \cite{related}.). 
At the LHC, the gluon fusion is the dominant Higgs boson production
 process and for light Higgs boson with mass $m_h < 150$ GeV, and 
 two photon decay mode of Higgs boson becomes 
 the primary discovery mode \cite{Djouadi} 
 nevertheless its branching ratio is ${\cal O}(10^{-3})$. 
The coupling between Higgs boson and these gauge bosons 
 are induced through quantum corrections at one-loop level even in the SM. 
Therefore, we can expect a sizable effect 
 from new particles if they contribute to the coupling at one-loop level. 
In a five dimensional GHU model, 
 we calculate one-loop diagrams with KK fermions 
 for the effective couplings between Higgs boson and the gauge bosons 
 (gluons and photons). 
If the KK mass scale is small enough, 
 we can see a sizable deviation from the SM couplings 
 and as a result, the number of signal events from Higgs production 
 at the LHC can be altered from the SM one. 
Interestingly, reflecting the special structure of Higgs sector in the GHU, 
 there is a clear qualitative difference from the UED scenario, 
 the signs of the effective couplings are opposite to those in the UED scenario.

\section{Model}
Let us consider a toy model 
 of five dimensional (5D) $SU(3)$ GHU with an orbifold $S^1/Z_2$ compactification, 
 in order to avoid unnecessary complications for our discussion. 
Although the predicted Weinberg angle in this toy model 
 is unrealistic, $\sin^{2} \theta_{W} = \frac{3}{4}$, 
 this does not affect our analysis. 
We introduce an $SU(3)$ triplet fermion as a matter field, 
 which is identified with top and bottom quarks 
 and their KK excited states, 
 although the top quark mass vanishes and the bottom quark mass 
 $m_{b} = M_{W}$ in this simple toy model. 
In our analysis, we will take into account a situation where a realistic top quark mass is realized 
 and the bottom quark contributions are negligible comparing to the top quark ones.  


The $SU(3)$ gauge symmetry is broken to $SU(2) \times U(1)$ 
 by the orbifolding on $S^1/Z_2$. 
The remaining gauge symmetry $SU(2) \times U(1)$ is supposed to be broken 
 by the vacuum expectation value (VEV) of the zero-mode of $A_5$, 
 the extra space component of the gauge field identified with the SM Higgs doublet. 
We do not address the origin of $SU(2) \times U(1)$ gauge symmetry 
 breaking and the resultant Higgs boson mass 
 in the one-loop effective Higgs potential, 
 which is highly model-dependent and out of our scope of this work.

The Lagrangian is simply given by 
\bea
{\cal L} = -\frac{1}{2} \mbox{Tr}  (F_{MN}F^{MN}) 
+ i\bar{\Psi}D\!\!\!\!/ \Psi. 
\label{lagrangian}
\eea
The periodic boundary conditions are imposed along $S^1$ for all fields. 
The non-trivial $Z_2$ parities are assigned for each field as follows, 
\bea 
\label{z2parity} 
A_{\mu}(y_i-y) &=& {\cal P} A_\mu(y_i+y) {\cal P}^\dag, \\
A_y(y_i-y) &=& -{\cal P} A_y(y_i+y) {\cal P}^\dag, \\
\Psi(y_i-y) &=& {\cal P} \gamma^5 \Psi(y_i+y)
\eea
where $\gamma^5 \psi_L=\psi_L$, ${\cal P}={\rm diag}(+,+,-)$ at fixed points $y_i=0, \pi R$.
By this $Z_2$ parity assignment, $SU(3)$ is explicitly broken to $SU(2) \times U(1)$. 
Higgs scalar field is identified with the off-diagonal block of zero mode $A_y^{(0)}$.

After the electroweak gauge symmetry breaking, 4D effective Lagrangian 
 among KK fermions, the SM gauge boson and Higgs boson ($h$) 
 defined as $h^0=(v+h)/\sqrt{2}$ can be derived from the term 
 ${\cal L}_{{\rm fermion}}  = i\bar{\Psi}D\!\!\!\!/ \Psi$ 
 in Eq.~(\ref{lagrangian}). 
Integrating over the fifth dimensional coordinate, 
 we obtain a relevant 4D effective Lagrangian in the mass eigenstate of nonzero KK modes:  
\bea
&&{\cal L}_{{\rm fermion}}^{(4D)}  
= \sum_{n=1}^{\infty} \left\{  (\bar{\psi}_1^{(n)}, 
\bar{\tilde{\psi}}_2^{(n)}, \bar{\tilde{\psi}}_3^{(n)}) 
\right. \nonumber \\ 
&& \left. \times \left(
\begin{array}{ccc}
i \gamma^{\mu} \partial_{\mu} - m_{n} & 0 & 0 \\
0 & i \gamma^{\mu} \partial_{\mu} 
 -\left( m_{+}^{(n)} + \frac{m}{v} h \right) & 0 \\
0& 0 &i \gamma^{\mu} \partial_{\mu} 
 -\left( m_{-}^{(n)} - \frac{m}{v} h \right)  
\end{array}
\right)
\left(
\begin{array}{c} 
\psi_1^{(n)} \\
\tilde{\psi}_2^{(n)} \\
\tilde{\psi}_3^{(n)}
\end{array}
\right) \right. \nonumber \\ 
&&   + \ \mbox{gauge interaction part} + \ \mbox{zero-mode part}.  
\label{4Deff}
\eea
where 
 $m = \frac{gv}{2} (= M_{W})$ is the bottom quark mass in this toy model,  
 $g = \frac{g_{5}}{\sqrt{2\pi R}}$ is the 4D gauge coupling. 
Note that the mass splitting $m_\pm^{(n)} \equiv m_{n} \pm m = \frac{n}{R} \pm m$ 
 occurs associated with a mixing 
 between the $SU(2)$ doublet component and singlet component. 
Note that the mass eigenstate for $m_\pm^{(n)}$ 
 has the Yukawa coupling $\mp m/v$, which is exactly the same 
 as the one for the zero mode. 
In UED, however, the KK mode mass spectrum and Yukawa couplings 
 are given by $M_n = \sqrt{m_n^2 + m_t^2}$ without mass splitting 
 and $-(m_t/v) \times (m_t/M_n)$, respectively.\cite{UED} 
Together with the mass splitting of KK modes, 
 this property is a general one realized in any GHU model 
 and leads to a clear qualitative difference of the GHU from the UED scenario, 
 as we will see. 

\section{Effective couplings between Higgs boson and gauge bosons}
Before calculating KK fermion contributions  
 to one-loop effective couplings between Higgs boson 
 and gauge bosons (gluons and photons), 
 it is instructive to recall the SM result. 
We parameterize the effective coupling between 
 Higgs boson and gluons or photons as 
\bea 
&&{\cal L}_{\rm eff} = C_g^{SM} \;  h \; G^{a\mu\nu}G^a_{\mu\nu}, \\
&&{\cal L}_{{\rm eff}} = C_{\gamma}^{SM} h F^{\mu\nu} F_{\mu\nu}, 
\eea
 where $G_{\mu\nu}^a (F_{\mu\nu})$ is a gluon (photon) field strength tensor. 
This coupling is generated by one-loop corrections 
 (triangle diagram) on which quarks are running. 
The top quark loop diagram gives the dominant contribution 
 and the coupling $C_g^{SM}$ and $C_{\gamma}^{SM}$ is described in the following form: 
\bea 
 C_g^{SM} &=& - \frac{m_t}{v} \times 
          \frac{\alpha_s F_{1/2}(4m_t^2/m_h^2)}{8\pi m_t} 
           \times \frac{1}{2} \simeq \frac{\alpha_s}{12\pi v},  \\           
 C_{\gamma}^{SM} &=& 
 -\frac{m_t}{v} \times  
 \frac{\alpha_{em} F_{1/2}(4m_t^2/m_h^2)}{8\pi m_t} 
 \times \frac{4}{3}  
  - \frac{m_W^2}{v} \times 
   \frac{\alpha_{em} F_1(4m_W^2/m_h^2)}{8 \pi m_W^2} \nonumber \\ 
 &\simeq& -\frac{47\alpha_{em}}{72\pi v}
\eea 
where the first (second) term in $C_\gamma^{SM}$ is KK top quark (KK W-boson) contributions, respectively,
 $\alpha_{s,em}$ is the fine structure constant of QCD, electromagnetic coupling, 
 the loop function $F_{1/2,1}(\tau)$ given by 
\bea
 F_{1/2}(\tau) &=& -2 \tau \left( 
   1+ \left( 1 - \tau \right) 
   [\sin^{-1}(1/\sqrt{\tau})]^2 \right) 
   \to -\frac{4}{3}~~{\rm for}~\tau \gg 1, \\
 F_1(\tau) &=& 
 2 + 3 \tau +3 \tau (2-\tau) [\sin^{-1}(1/\sqrt{\tau})]^2 \to  7 \; \mbox{for} \; \tau \gg 1. 
\label{loopfunc}
\eea
It is well-known that in the top quark decoupling limit $m_t \gg m_h$,  
 $F_{1/2,1}$  becomes a constant and 
 the resultant effective coupling becomes independent of $m_t$, $m_W$ and $m_h$.

\begin{figure}[t]
\begin{center}
\psfig{file=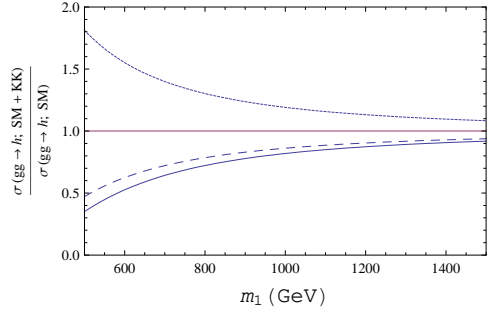,width=2.5in}
\end{center}
\caption{
The ratio of the Higgs boson production cross sections 
 in the GHU and in the SM as a function of the KK mode mass $m_1$. 
The solid (dashed) line corresponds to 
 the result including the (half-)periodic fermion contributions, respectively. 
As a reference, the result in the UED scenario 
 with top quark KK modes is also shown (dotted line). 
We have taken $m_h=120$ GeV. 
}
\label{fig1}
\end{figure}

Calculations of KK mode contributions are completely analogous 
 to the top loop correction. 
The structure described in our toy model is common 
 in any GHU model, 
 we will have KK modes of top quark 
 with mass eigenvalue $m_{\pm}^{(n)}= m_n \pm m_t$ 
 and Yukawa couplings $ \mp m_t/v$, respectively. 
The KK mode contributions are found to be  
\bea
{\cal L}_{{\rm eff}} &=& C_g^{KK(GH)} \; 
  h \; G^{a\mu\nu} G^a_{\mu\nu} , \nonumber \\ 
 C_g^{KK(GH)} &=& -\sum_{n=1}^\infty 
 \left[
 \frac{m_t}{v} \times 
 \frac{\alpha_s F_{1/2}(4 m_{+}^{(n)2}/m_h^2)}
 {8\pi m_{+}^{(n)}} \times \frac{1}{2} \right] + (m_{+}^{(n)} \leftrightarrow m_{-}^{(n)})\nonumber \\
&\simeq & 
 \frac{m_t \alpha_s}{12 \pi v}
 \sum_{n=1}^\infty 
\left[ \frac{1}{m_{+}^{(n)}} - \frac{1}{m_{-}^{(n)}} \right] 
 \simeq  - \frac{\alpha_s}{6 \pi v}
 \sum_{n=1}^\infty \frac{m_t^2}{m_n^2}  
\eea
where we have taken the limit $m_h^2,~m_t^2 \ll m_n^2$ to simplify the results. 
Note that this result is finite due to the cancellation 
 between two divergent corrections with opposite signs. 
Also, note that the KK mode contribution is subtractive 
 against the top quark contribution in the SM.  
The results are depicted in Fig.~\ref{fig1} 
 as a function of the mass of the lightest KK mode (diagonal) mass eigenvalue ($m_1$). 
For the bulk fermion with the (half-)periodic boundary condition $m_1 = 1/R (1/(2R))$. 
In this analysis, we take $m_h=120$ GeV.  
The result is not sensitive to the Higgs boson mass if $m_h < 2 m_t$. 
For reference, 
 the result in the UED scenario \cite{UED} is also shown, 
 for which only the periodic fermion has been considered. 
The KK fermion contribution is subtractive 
 and the Higgs production cross section is reduced in the GHU, 
 while it is increased in the UED scenario. 
This is a crucial point to distinguish the GHU from the UED scenario. 

The contribution of top quark KK modes 
 to the effective coupling between Higgs boson and photons 
 are calculated similarly. 
\bea
{\cal L}_{{\rm eff}} &=& C_{\rm \gamma}^{KK(GH)} \; 
  h \; F^{\mu\nu} F_{\mu\nu} , \nonumber \\ 
 C_{{\rm \gamma}}^{KK(GH)} &=& -\sum_{n=1}^\infty 
 \left[
 \frac{m_t}{v} \times 
 \frac{\alpha_{em} F_{1/2}(4 m_{+}^{(n)2}/m_h^2)}
 {8\pi m_{+}^{(n)}} \times \frac{4}{3} \right] +(m_{+}^{(n)} \leftrightarrow m_{-}^{(n)}) \nonumber \\
&\simeq & 
 \frac{2 m_t \alpha_{em}}{9 \pi v}
 \sum_{n=1}^\infty 
\left[ \frac{1}{m_{+}^{(n)}} - \frac{1}{m_{-}^{(n)}} \right] 
 \simeq  - \frac{4 \alpha_{em}}{9 \pi v}
 \sum_{n=1}^\infty \frac{m_t^2}{m_n^2}.  
\label{gghGH}
\eea
 
For the effective coupling with photons, 
 in addition to the KK fermion contributions,  
 we also have the KK W-boson loop corrections as in the SM. 
However, we neglect such contributions 
 compared to those from the KK top quark ones by the following plausible reasons. 
\begin{itemlist}
\item KK top (KK W-boson) contributions are decoupling effects, 
         and are proportional to mass squared of top (W-boson), respectively. 
         This indicates that KK top quark contributions are likely to be dominant. 
\item In the GHU model on the flat space, 
         a large dimensional representation in which the SM top quark is embedded 
         must be introduced to reproduce a realistic top Yukawa coupling \cite{CCP}.  
         Therefore, the effective 4D theory includes extra vector-like top-like quarks and its KK modes. 
         Thus, KK top quark contributions are enhanced by a number of extra top-like quarks. 
\item In some GHU models, 
        bulk top-like quarks with the half-periodic boundary condition 
        are often introduced to realize the correct electroweak symmetry 
        breaking and a viable Higgs boson mass. 
        The lowest KK mass of the half-periodic fermions is half of the lowest KK mass of periodic ones, 
        so that their loop contributions can dominate over those by periodic KK mode fields. 
\end{itemlist} 
After these considerations, 
 we see that the KK mode contributions to two photon decay is the same sign as those of the SM. 

\section{Effects on Higgs boson search at LHC} 
As we have shown, the KK mode loop contribution to 
 the effective coupling between Higgs boson and 
 gluons (photons) is subtractive (slightly constructive) to 
 the top quark loop contribution in the SM. 
This fact leads to remarkable effects on Higgs boson search at the LHC.  
Since the main production process of Higgs boson at the LHC is through gluon fusion, 
and the primary discovery mode of Higgs boson is 
 its two photon decay channel if Higgs boson is light $m_h < 150$ GeV. 
Therefore, the deviations of the effective coupling 
 between Higgs boson and gluons or photons from the SM one 
 give important effects on the Higgs boson production and 
the number of two photon events from Higgs boson decay. 

\begin{figure}[h]
\begin{center}
\psfig{file=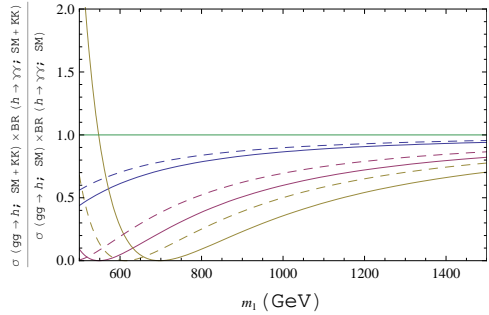,width=2.5in}
\end{center}
\caption{
The ratio of the number of two photon events 
 in the GHU scenario  
 to those in the SM as a function of the lowest KK mass $m_1$. 
The solid (dashed) lines represent the results 
 including the $n_t$ (half-)periodic KK fermion contributions. 
 $n_t=1,3,5$ are shown from the top to the bottom at $m_1=1500$ GeV. 
Higgs mass is taken to be 120 GeV.  
}
\label{fig4}
\end{figure}

We show the ratio of the number of two photon events  
 from Higgs decay produced through gluon fusion at the LHC. 
As a good approximation, 
 this ratio is described as 
\bea 
&& \frac{
   \sigma(gg \to h;~{\rm SM+KK}) 
      \times BR(h \to \gamma \gamma;~{\rm SM+KK})}
   {\sigma(gg \to h;~{\rm SM}) 
      \times BR(h \to \gamma \gamma;~{\rm SM})} \nonumber \\ 
&\simeq& 
 \left(1+ \frac{C_g^{KK(GH)}}{C_g^{SM}} \right)^2 
 \left(1+ \frac{C_\gamma^{KK(GH)}}{C_\gamma^{SM}} \right)^2 .
\eea 
where $\sigma$ is Higgs boson production cross section, 
 BR denotes the branching ratio of two photon decay of Higgs boson.  
Fig.~\ref{fig4} shows the results 
 for the periodic and half-periodic KK modes 
 as a function of $m_1$ for the case of $n_t=$1, 3 and 5 extra top-like quark fermions. 
Even for $m_1=1$ TeV and $n_t=1$, 
 the deviation is sizable $\simeq$14\%. 
When $m_1$ is small and $n_t$ is large, 
 the new physics contribution can dominate. 

\section{Summary}


We have calculated the one-loop KK fermion contributions to 
 the Higgs effective couplings between Higgs boson and gluons or photons 
 and found them to be finite. 
This finiteness is achieved by a non-trivial cancellations 
 between two KK mass eigenstates, 
 although each contribution is divergent. 
The overall sign of the contributions is opposite 
 compared to the SM result by top quark loop corrections 
 and the similar result in the UED scenario. 
Therefore, this feature is a clue to distinguish the GHU from the UED scenario. 
Our analysis have shown that even with the KK mode mass 
 is around 1 TeV, the KK mode loop corrections provide 
 ${\cal O}$(10\%) deviations from the SM results 
 in Higgs boson phenomenology at the LHC.  
In a realistic GHU model, 
 some extra top-like quarks would be introduced  
 to reproduce the top Yukawa coupling in the SM. 
In such a case, 
 the KK mode contributions are enhanced and 
 the signal events of Higgs boson production at the LHC 
 are quite different from those in the SM.

\section*{Acknowledgments}
The work of authors was supported 
in part by the Grant-in-Aid for Scientific Research 
of the Ministry of Education, Science and Culture, No.18204024 and No.20025005.



\end{document}